\documentclass[prl,twocolumn]{revtex4}

\usepackage{graphicx}
\usepackage{dcolumn}
\usepackage{amsmath}
\usepackage {epsfig}

\makeatother

\begin{document}

\title[Short Title]{ Color superconductivity, BPS 
strings and monopole confinement in \protect\( \mathcal{N}=2\protect \) 
and \protect\( \mathcal{N}=4\protect \)
super Yang-Mills theories }

\author{Marco A. C. Kneipp} 
\affiliation{Universidade do Estado do Rio de Janeiro(UERJ), 
Dept. de F\'\i sica Te\' orica,\\
Rua S\~ao Francisco Xavier, 524,\\
20550-130 Rio de Janeiro, Brazil \\
and Centro Brasileiro de Pesquisas F\' \i sicas (CBPF), \\ 
Coordena\c c\~ ao de Teoria de Campos e Part\'\i culas (CCP),\\
Rua Dr. Xavier Sigaud, 150\\
22290-180 Rio de Janeiro, Brazil. }

\begin{abstract}
We review some recent developments on BPS string solutions and monopole confinement
in the Higgs or (color) superconducting phase of  deformed \( \mathcal{N}=2 \)
and \( \mathcal{N}=4 \) super Yang-Mills theories. In particular, the monopole
magnetic fluxes are shown to be always integer linear combinations of string
fluxes. Moreover, a bound for the threshold length of the string breaking is
obtained. When the gauge group \( SU(N) \) is broken to \( Z_{N} \), the BPS
string tension satisfies the Casimir scaling law. Furthermore in the \( SU(3) \)
case the string solutions are such that they allow the formation of a confining
system with three monopoles.
\end{abstract}

\maketitle

\section{Introduction}

It has long been believed that particle confinement at the strong coupling regime
should be a phenomenon dual to monopole confinement in a (color) superconductor
in weak coupling. Therefore, the study of monopole confinement in weak coupling
may shed some light on particle confinement. Since dualities are better understood
for supersymmetric theories, it is interesting to analyze monopole confinement
in these theories. 

We shall review the results in \cite{kb, k, k2003} where we analyzed monopole
confinement in non-Abelian Yang-Mills-Higgs theories at the weak coupling regime
with two symmetry breaking. In the first symmetry breaking the theory is in
the Coulomb phase with solitonic monopole solutions which may fill representations
of a non-Abelian group. Then, in the second symmetry breaking, the theory is
in the Higgs or (color) superconducting phase with strings or flux tubes. We
show explicitly that in these theories always the magnetic fluxes of the monopoles
are integer multiple of the strings fluxes. The first symmetry breaking is due
to the expectation value of a complex scalar \( \phi _{3} \) in the adjoint
representation. Then, the second symmetry breaking is due to two complex scalars
\( \phi _{1} \) and \( \phi _{2} \) in complex conjugated representations.
In order to exist topological string solutions, two possible representations
are considered: in \cite{k2003} \( \phi _{1} \) and \( \phi _{2} \) are in
the adjoint representation. On the other hand, in \cite{kb, k} \( \phi _{1} \)
is in the representation \( R^{\textrm{sym}}_{k\lambda _{\phi }} \) which is
the symmetric part of the direct product of \( k \) fundamental representations
\( R_{\lambda _{\phi }} \), with \( k\geq 2 \), and in \( \phi _{2} \) in
the complex conjugated representation. In particular, if \( k=2 \), this is
exactly the same representation as that of a diquark condensate, where by quark
we mean a fermion in a fundamental representation \( R_{\lambda _{\phi }} \)
of the gauge group \( G \). We have chosen the potential to be the bosonic
part of \( \mathcal{N}=4 \) or \( \mathcal{N}=2 \) super Yang-Mill (SYM) theories
with some deformation mass terms. These potentials appear naturally from the
BPS string conditions. One of the main difference between the two analyzed representations
for \( \phi _{1} \) and \( \phi _{2} \) mentioned above, is the following:
when one considers the representation \( R^{\textrm{sym}}_{k\lambda _{\phi }} \),
during the second symmetry breaking, just a \( U(1) \) factor inside \( G \)
is broken to a discrete subgroup, similarly to what happens in a superconductor,
and it produces a monopole-antimonopole confinement. On the other hand, when
\( \phi _{1} \) and \( \phi _{2} \) are in the adjoint representation, the
full group \( G \) is broken to a discrete subgroup, which produce a color
superconductor.  For \( G=SU(3) \) we showed that this kind of breaking produces
a confining system with three different monopoles, besides the monopoles-antimonopole
system. We also showed explicitly for \( G=SU(N) \) that the BPS string tensions
satisfy the Casimir scaling law. In \cite{witten,  strassler} was pointed out
that this deformend \( \mathcal{N}=4 \) SYM theories should have a weakly coupled
Higgs phase with magnetic flux tubes and this phase should be dual to a strongly
coupled confining phase in the dual theory. One of our aims was to analyze many
properties of these magnetic flux tubes. 

When the scalars \( \phi _{1} \) and \( \phi _{2} \) are in the same representation
as that of a diquark condensate, one could think of \( \phi _{1} \) and \( \phi _{2} \)
as being themselves diquark condensates. In this case, we would have a situation
quite similar to the one in an ordinary superconductor, described by the Abelian-Higgs
theory with the scalar being a Cooper pair. If \( G=SU(N) \), the scalar(s)
in the adjoint representation could also be thought to be interpreted as quark-antiquark
condensate(s). However it is important to note that all the results described
here do not depend if the scalars are condensates or not. For \( G=SU(3) \),
these two kinds of condensates are the color sextet and octet. They are expected
to exist in the color superconducting phase of (dense) QCD at the weak coupling
\cite{colorsupercondunting, colorsuper2}. The effective theory describing these
condensates are not well known. One could think that the theory considered here
when the gauge group is \( G=SU(3) \), as been a toy model for an effective
theory of these condensates. Then, one conclude that an effective theory for
these condensates could have monopoles, flux tubes and monopole confinement,
depending on the form of the potential. In the dual theory, one might conjecture
that these scalars could be monopole-monopole and monopole-antimonopole condensates.

\section{Deformed \protect\( \mathcal{N}=2\protect \) Super Yang-Mills theories}

As is well known, the Abelian-Higgs in the broken phase is an effective theory
for a superconductor with the complex scalar field \( \phi  \) being interpreted
as a electron pair condensate. In this theory since the \( U(1) \) gauge group
is broken to a discrete subgroup there are topological flux tubes or string
solutions with string tensions \( T \) satisfying 
\begin{equation}
\label{2.3a}
T\geq \frac{1}{2}q_{\phi }a^{2}\left| \Phi _{\textrm{st}}\right| \, ,
\end{equation}
 where 
\begin{equation}
\label{fluxo}
\Phi _{\textrm{st}}\equiv \int d^{2}x\, B_{3}=\frac{2\pi }{q_{\phi }}n\, ,\, \, \, n\in Z
\end{equation}
is the string magnetic flux and \( q_{\phi }=2e \) is the electric charge of
\( \phi  \). The lower bound in (\ref{2.3a}) is attained by the BPS string.
If one puts a (Dirac) monopole and antimonopole in a superconductor, their magnetic
lines could not spread over space but must rather form a string which gives
rise to a confining potential between the monopoles. This idea only makes sense
since the (Dirac) monopole magnetic flux is \( \Phi _{\textrm{mon}}=g=2\pi /e\, , \)
which is an integer multiple of the string's magnetic flux quantization condition
(\ref{fluxo}), allowing one to attach to the monopole two strings with \( n=1 \).
Then, using the electromagnetic duality of Maxwell theory one could map this
monopole confining system in the weak coupling regime to an electric charge
confining system in the strong coupling. 

Let us generalize some of these ideas to a non-Abelian theory. Let us consider
an arbitrary gauge group \( G \), without \( U(1) \) factors and such that
\( \Pi _{0}(G)=0=\Pi _{1}(G) \), like for example \( G=SU(N) \). In \cite{kb, k},
we considered the Lagrangian
\begin{eqnarray}
L & = & -\frac{1}{4}G^{\mu \nu }_{a}G_{a\mu \nu }+\frac{1}{2}\left( D_{\mu }\phi _{3}\right) ^{*}_{a}\left( D^{\mu }\phi _{3}\right) _{a}+\nonumber \\
 &  & +\frac{1}{2}\sum _{i=1}^{2}\left( D_{\mu }\phi _{i}^{\dagger }\right) \left( D^{\mu }\phi _{i}\right) -V(\phi )\label{1.2} 
\end{eqnarray}
 with potential given by
\[
V(\phi )=\frac{1}{2}\left( \sum _{p=1}^{3}\left( d_{a}^{p}\right) ^{2}+\sum _{m=1}^{2}F_{m}^{\dagger }F_{m}\right) \]
 where 
\begin{eqnarray}
d^{3}_{a} & = & \frac{e}{2}\left( \phi ^{*}_{3b}if_{bca}\phi _{3c}+\phi _{m}^{\dagger }\sigma ^{3}_{mn}T_{a}\phi _{m}-m\textrm{Re}\left( \phi _{3a}\right) \right) ,\nonumber \label{4.2a} \\
d_{a}^{p} & = & \frac{e}{2}\left( \phi _{m}^{\dagger }\sigma ^{p}_{mn}T_{a}\phi _{n}\right) ,\, \, \, \, p=1,2,\nonumber \label{4.2b} \\
F_{1} & = & e\left( \phi _{3a}^{\dagger }T_{a}-\frac{\mu }{e}\right) \phi _{1},\nonumber \label{4.2c} \\
F_{2} & = & e\left( \phi _{3a}T_{a}-\frac{\mu }{e}\right) \phi _{2},\nonumber \label{4.2d} 
\end{eqnarray}
 with \( \sigma ^{p} \) being the Pauli matrices and \( T_{a} \) being the
generators of \( G \). This potential is the bosonic part of \( \mathcal{N}=2 \)
super Yang-Mills with one flavor and a breaking mass term. The scalar \( \phi _{3a} \)
in the adjoint representation belongs to the vector supermultiplet and the scalars
\( \phi _{1} \) and \( \phi _{2} \) belong to a massive hypermultiplet. The
real parameter \( \mu  \) gives a bare mass to \( \phi _{1} \) and \( \phi _{2} \)
and \( m \) gives a bare mass to the real part of \( \phi _{3} \) and therefore
breaks \( \mathcal{N}=2 \) supersymmetry to \( \mathcal{N}=0 \). In \cite{kb},
we started with a generic potential and have shown that in order to obtain the
BPS string conditions, the potential is almost constrained to have this form.
We shall consider the theory in the weak coupling regime, and therefore we shall
not consider the quantum corrections to the potential.

\section{Phases of the theory}

Let us review very quickly some of the Lie algebra conventions adopted. The
Lie algebra generators satisfy the commutation relations
\begin{eqnarray*}
\left[ H_{i},H_{j}\right]  & = & 0\\
\left[ H_{j},E_{\alpha }\right]  & = & \alpha ^{j}E_{\alpha },\\
\left[ E_{\alpha ,}E_{-\alpha }\right]  & = & \frac{2\alpha \cdot H}{\alpha ^{2}},
\end{eqnarray*}
where the upper index in \( \alpha ^{j} \) means the \( j \) component of
the root \( \alpha  \). Let us denote by \( \alpha _{i} \) the simple roots
and \( \lambda _{i} \) the fundamental weights which satisfy the relation
\[
\frac{2\lambda _{i}\cdot \alpha _{j}}{\alpha _{i}^{2}}=\delta _{ij}.\]
 The weights states \( \left| \omega \right\rangle  \) of a representation
satisfy 
\[
v\cdot H\left| \omega \right\rangle =v\cdot \omega \left| \omega \right\rangle .\]
 As mentioned in the introduction, in \cite{kb, k}, we considered \( \phi _{1} \)
in \( R^{\textrm{sym}}_{k\lambda _{\phi }} \), the symmetric part of the direct
product of \( k \) fundamental representations \( R^{\textrm{sym}}_{\lambda _{\phi }} \),
where \( k\geq 2 \) and \( \lambda _{\phi } \) is an arbitrary fundamental
weight. This representation possess in particular the weight state \( \left| k\lambda _{\phi }\right\rangle  \),
which will be responsible for one of the symmetry breakings as we shall see.

Returning to our physical problem. The vacua must be solutions of \( V(\phi )=0 \)
which is equivalent to 
\begin{equation}
\label{2.4}
d^{p}=0=F_{m}\, .
\end{equation}
In order to the topological string solutions to exist, we look for vacuum solutions
of the form 
\begin{eqnarray}
\phi ^{\textrm{vac}}_{1} & = & a\left| k\lambda _{\phi }\right\rangle ,\nonumber \label{2.5a} \\
\phi _{2}^{\textrm{vac}} & = & 0\label{2.5b} \\
\phi _{3}^{\textrm{vac}} & = & b\lambda _{\phi }\cdot H,\nonumber   \\
W^{\textrm{vac}}_{\mu } & = & 0\, ,\nonumber 
\end{eqnarray}
where \( a \) is a complex constant and \( b \) is real. As explained in detail
in \cite{OT, kb, k}, the above vacuum configuration produce a symmetry breaking
\begin{eqnarray}
G & \rightarrow  & G_{\phi _{3}}\equiv [K\times U(1)]/Z_{l}\, \rightarrow \nonumber \\
 & \rightarrow  & G_{\phi _{1}}\equiv [K\times Z_{kl}]/Z_{l}\label{3.1r} 
\end{eqnarray}
 where \( K \) is a subgroup of \( G \) and \( Z_{l} \) is a discrete subgroup
of \( U(1) \) and \( K \). In the particular case \( G=SU(N) \) and \( \lambda _{\phi }=\lambda _{1} \),
the fundamental weight of the \( N \) dimensional representation, we have the
symmetry breaking
\begin{eqnarray}
SU(N) & \rightarrow  & G_{\phi _{3}}\equiv [SU(N-1)\times U(1)]/Z_{N-1}\, \rightarrow \nonumber \\
 & \rightarrow  & G_{\phi _{1}}\equiv [SU(N)\times Z_{k(N-1)}]/Z_{N-1}\label{3.2r} 
\end{eqnarray}
 The first symmetry breaking is due to \( \phi _{3}^{\textrm{vac}} \), with
\( b\neq 0, \) and the second is due to \( \phi _{1}^{\textrm{vac}} \), with
\( a\neq 0 \). 

From the vacuum equations (\ref{2.4}) one can conclude that 
\begin{eqnarray*}
|a|^{2} & = & \frac{mb}{k}\, ,\\
\left( kb\lambda _{\phi }^{2}-\frac{\mu }{e}\right) a & = & 0\, .
\end{eqnarray*}
 There are three possibilities:

\begin{description}
\item [(i)]If \( m\mu <0\, \, \Rightarrow \, \, a=0=b \) and the gauge group \( G \)
remains unbroken.
\item [(ii)]If \( m=0,\, \mu \neq 0\, \, \Rightarrow \, \, a=0 \) and \( b \) can
be any constant. In this case, \( \phi _{3}^{\textrm{vac}} \) produces the
first symmetry breaking in (\ref{3.1r}) or (\ref{3.2r}) which corresponds
to the Coulomb phase.
\item [(iii)]If \( m\mu >0\, \Rightarrow  \)
\begin{equation}
\label{2.7b}
|a|^{2}=\frac{m\mu }{k^{2}e\lambda _{\phi }^{2}}\, \, ,\, \, \, \, \, \, \, \, b=\frac{\mu }{ke\lambda _{\phi }^{2}}
\end{equation}
 and it happens the second symmetry breaking, which corresponds to the Higgs
or superconducting phase.
\end{description}
Let us analyze each of these phases.

\section{Coulomb phase}

This phase occurs when \( G \) is broken to \( G_{\phi _{3}} \). The \( U(1) \)
factor in \( G_{\phi _{3}} \) is generated by \( \phi _{3}^{\textrm{vac}} \).
As we have seem, that symmetry breaking can happen only when \( m=0 \) and
therefore the \( \mathcal{N}=2 \) symmetry is restored since \( m \) was a
supersymmetry breaking parameter. In this phase, since \( \Pi _{2}(G/G_{\phi _{3}})=Z \),
there exist magnetic monopoles. The stable or fundamental BPS monopoles are
those with lowest magnetic charge \cite{eweinberg}. These fundamental monopoles,
are believed to fill representations of the gauge subgroup \( K^{\textrm{v}} \)
\cite{12, HolloDoFraMACK}. The magnetic charges of monopoles for a general
symmetry breaking has been obtained long time ago in \cite{bais, eweinberg}.
In particular for the first symmetry breaking in (\ref{3.1r}) or (\ref{3.2r}),
the magnetic charge for the fundamental monopoles, can be written as \cite{k}
\begin{equation}
\label{3.6}
g\equiv \frac{1}{|\phi _{3}^{\textrm{vac}}|}\int dS_{i}\textrm{Re}\left( \phi _{3}^{a}\right) B_{i}^{a}=\frac{2\pi }{e|\lambda _{\phi }|}=\frac{2\pi k}{q_{\phi }}
\end{equation}
where the integral is taking over the closed surface surrounding the monopole,
\( B_{i}^{a}\equiv -\epsilon _{ijk}G_{jk}^{a}/2 \) are the non-Abelian magnetic
fields and 
\begin{equation}
\label{4.1}
q_{\phi }=ek|\lambda _{\phi }|\, .
\end{equation}
is electric charge of \( \phi _{1}^{\textrm{vac}} \) \cite{k}. In Eq. (\ref{3.6})
appears the real part of \( \phi _{3a} \) since it is the real part of the vacuum 
configuration \( \phi _{3}^{\textrm{vac}} \)
which is responsible for  the first symmetry breaking. 

These monopoles fill supermultiplets of \( \mathcal{N}=2 \) supersymmetry and
satisfy the mass formula 
\begin{equation}
\label{3.6a}
M_{\textrm{mon}}=|\phi _{3}^{\textrm{vac}}||g|\, .
\end{equation}
 
In particular, for the symmetry breaking 
\[
SU(N+1)\, \rightarrow \, [SU(N)\times U(1)]/Z_{N},\]
since \( |\lambda _{\phi }|=|\lambda _{1}|=\sqrt{N/(N+1)} \), then from Eq.
(\ref{3.6}), it results 
\footnote{Note that for \( G=SU(2) \), this result is obtained 
considering that \( |\lambda _{1}|=1/\sqrt{2}. \)
However, for this group it is usually adopted the convention \( |\lambda _{1}|=1/2 \),
which results the standard charge \( g=4\pi /e \) for the stable monopole.
} 
\[
g=\frac{2\pi }{e}\sqrt{\frac{N+1}{N}}.\]
In this case, the fundamental monopoles are expected to fill the \( N \) dimensional
representation of \( SU(N) \) \cite{12, HolloDoFraMACK}. As was pointed out
long time ago \cite{corriganolive}, due to the monopole solutions for a symmetry 
breaking of this type, the
\( U(1) \) electric charge \( q_{c} \) of a particle in the fundamental representation
of the unbroken group \( SU(N) \) must satisfy the quantization condition
\[
q_{c}=\frac{m}{N}q_{0},\, \, \, \, m\in Z\]
where \( q_{0} \) is the \( U(1) \) electric charge of a \( SU(N) \) singlet.
That is a generalization of Dirac quantization condition which, for the case
of \( N=3 \), gives the right electric charge quantization condition for the
quarks.

\section{Higgs or superconducting phase}

\subsection{The BPS \protect\( Z_{k}\protect \)-string solutions}

This phase occurs when \( G \) is broken to \( G_{\phi _{1}} \). Moreover,
since \( m\neq 0 \), \( \mathcal{N}=2 \) supersymmetry is broken to \( \mathcal{N}=0 \).
In this phase, the \( U(1) \) factor in \( G_{\phi _{3}} \) is broken to the
discrete subgroup \( Z_{k} \) and, like in the Abelian-Higgs theory, the magnetic
flux lines associated to this \( U(1) \) factor cannot spread over space. However,
since \( G \) is broken in such a way that \( \Pi _{1}(G/G_{\phi _{1}})=Z_{k} \),
these flux lines may form topological \( Z_{k}- \)strings. In \cite{kb}, a
\( Z_{k} \) string ansatz was constructed, associated to each of the \( (k-1) \)
non trivial group elements of the discrete group \( Z_{k} \). We have also
obtained the BPS string conditions. Putting the ansatz into these BPS conditions
we obtained that the functions which appear in the ansatz must satisfy exactly
the same differential equations with same boundary conditions as for the BPS
string in the Abelian-Higgs theory. The existence of non trivial solutions for
these differential equations has been proven by Taubes\cite{taubes}.

\subsection{\protect\( Z_{k}\protect \)-string magnetic flux, monopole confinement and
the string tension}

In this phase, the monopole's magnetic lines associated to the broken \( U(1) \)
factor can no longer spread radially over space. However, these \( U(1) \)
could form flux tubes and the monopole get confined. In order for that to happen,
the monopole flux \( \Phi _{\textrm{mon}} \) in this \( U(1) \) direction,
which is equal to the magnetic charge (\ref{3.6}), must be an integer multiple
of the string fluxes \( \Phi _{\textrm{st}} \) in this \( U(1) \) direction.
We define
\begin{equation}
\label{4.3}
\Phi _{\textrm{st}}\equiv \frac{1}{|\phi _{3}^{\textrm{vac}}|}\int d^{2}x\, \textrm{Re}(\phi ^{a}_{3})B^{a}_{3}
\end{equation}
 similarly to the monopole magnetic flux definition (\ref{3.6}), but with surface
integral taken over the plane perpendicular to the string. We obtained for our
BPS \( Z_{k} \) string solutions that 
\begin{equation}
\label{4.4c}
\Phi _{\textrm{st}}=\frac{2\pi n}{q_{\phi }},\, \, \, \, n\in Z_{k}
\end{equation}
where each value \( n \) is associated to a \( Z_{k} \) group elements used
to construct the \( (k-1) \) solutions.

Therefore we can conclude that \( \Phi _{\textrm{mono}} \) can be equal for
example to \( k \) times \( \Phi _{\textrm{st}} \) for \( n=1 \). This can
be interpreted that for one monopole we could attach \( k \) \( Z_{k} \)-strings
with \( n=1 \). That is consistent with the fact the set \( k \) \( Z_{k}- \)strings
with \( n=1 \) belongs to the trivial sector of \( \Pi _{1}(G/G_{\phi _{1}}) \)
and therefore can terminate at some point. However, since it has a non-vanishing
magnetic flux it must terminate in a magnetic source, i.e., a monopole. It is
important to stress the fact that being in the trivial topological sector does
not mean that this set of strings has total vanishing flux. For the particular
case \( G=SU(2) \) and \( k=2 \), the field \( \phi _{1} \) is in the three
dimensional representation which is the adjoint of \( SU(2) \). Then we can
see that all these results are consistent with some well-known results for the
\( Z_{2} \) string of \( SU(2) \) Yang-Mills-Higgs theory, as explained in
\cite{hk review, Vilenkin}. In this theory there are at least two complex scalars
in the adjoint representation which produce the symmetry breakings \( SU(2)\, \rightarrow \, U(1)\, \rightarrow \, Z_{2} \).
In the Higgs phase, the stable \( Z_{2} \) string has flux \( 2\pi /e \).
In this phase, two strings get attached to a 't Hooft-Polyakov monopole with
magnetic charge \( g=4\pi /e \), and can produce the monopole-antimonopole
confinement.

We have shown that the string tension must satisfy the bound \cite{kb} 
\begin{equation}
\label{4.4d}
T\geq \frac{me}{2}\left| \phi _{3}^{\textrm{vac}}\right| \left| \Phi _{\textrm{st}}\right| =\frac{1}{2}q_{\phi }|a|^{2}|\Phi _{\textrm{st}}|,
\end{equation}
where \( |a| \) given by Eq. (\ref{2.7b}) is the modulus of \( \phi _{1}^{\textrm{vac}} \),
which produces the second symmetry breaking. That result is very similar to
the \( U(1) \) result given by Eq. (\ref{2.3a}). The string tension bound
hold for the BPS string. Since the tension is constant, it produces a confining
potential between monopoles increasing linearly with their distance. From string
tension bound one can obtain easily that the threshold length \( d^{\textrm{th}} \)
for the set of strings to break producing a new monopole-antimonopole pair,
with masses (\ref{3.6a}), satisfies the bound\cite{k}
\[
d^{\textrm{th}}\leq \frac{4}{me}\, .\]

It is interesting to note that, unlike the Abelian-Higgs theory, in our theory
the bare mass \( \mu  \) of \( \phi _{1} \) and \( \phi _{2} \) is not required
to satisfy \( \mu ^{2}<0 \) in order to happen the spontaneous symmetry breaking.
Therefore, since one could interpret \( \phi _{1} \)and \( \phi _{2} \) as
monopole condensates (when \( k=2 \)) in the dual theory, the monopole mass
do not need to satisfy the problematic condition \( M^{2}_{\textrm{mon}}<0 \)
mentioned by 't Hooft \cite{thooftreview}. The same thing happens in the theory
where all the scalar are in the adjoint, analyzed in the next sections.

\section{Deformed \protect\( \mathcal{N}=4\protect \) (or \protect\( \mathcal{N}=2^{*}\protect \))
super Yang-Mills theories}

Let us now analyze the monopole confinement in the theory with three complex
scalars \( \phi _{s} \), \( s=1,2,3 \), in the adjoint, as considered in \cite{k2003}.
Once more we shall consider a gauge group \( G \) without \( U(1) \) factors
and such that \( \Pi _{0}(G)=0=\Pi _{1}(G) \). Let us consider the Lagrangian
\[
L=-\frac{1}{4}G_{a\mu \nu }G_{a}^{\mu \nu }+\frac{1}{2}\left( D_{\mu }\phi _{s}^{*}\right) _{a}\left( D^{\mu }\phi _{s}\right) _{a}-V(\phi ).\]
 We shall consider the potential
\begin{equation}
\label{2.5}
V(\phi )=\frac{1}{2}\left[ \left( d_{a}\right) ^{2}+f_{sa}^{\dagger }f_{sa}\right] 
\end{equation}
where 
\begin{equation}
\label{2.1}
d_{a}\equiv \frac{e}{2}\left( \phi ^{*}_{sb}if_{abc}\phi _{sc}-m\textrm{Re}\left( \phi _{3a}\right) \right) ,
\end{equation}
 and 
\begin{eqnarray}
f_{1} & \equiv  & \frac{1}{2}\left( e\left[ \phi _{3},\phi _{1}\right] -\mu \phi _{1}\right) \, ,\nonumber \\
f_{2} & \equiv  & \frac{1}{2}\left( e\left[ \phi _{3},\phi _{2}\right] +\mu \phi _{2}\right) \, ,\label{2.6} \\
f_{3} & \equiv  & \frac{1}{2}\left( e\left[ \phi _{1},\phi _{2}\right] -\mu _{3}\phi _{3}\right) \, .\nonumber 
\end{eqnarray}
 This is the potential of the bosonic part of \( \mathcal{N}=4 \) super Yang-Mills
(SYM) theory with some mass term deformations which break completely supersymmetry.
If we set \( m=0 \), \( \mathcal{N}=1 \) supersymmetry is restored and we
obtain the potential considered in \cite{witten}. If further \( \mu _{3}=0 \)
we recover the potential of \( \mathcal{N}=2 \) with a massive hypermultiplet
in the adjoint representation. Finally, if also \( \mu =0 \), we obtain \( \mathcal{N}=4 \).
As usual, we shall denote by \( \mathcal{N}=2^{*} \), \( \mathcal{N}=1^{*} \)
and \( \mathcal{N}=0^{*} \) to the theories which are obtained by adding deformation
mass terms to \( \mathcal{N}=4 \) SYM theory.

\section{Phases of the theory}

The vacua of the theory are solutions of
\begin{equation}
\label{3.1}
G_{\mu \nu }=D_{\mu }\phi _{s}=V(\phi )=0\, .
\end{equation}
The condition \( V(\phi _{s})=0 \) is equivalent to 
\begin{equation}
\label{3.2}
d_{a}=0=f_{sa}\, .
\end{equation}
 We are looking for vacuum solutions which produce the symmetry breaking 
\[
G\, \rightarrow \, U(1)^{r}\, \rightarrow \, C_{G,}\]
where \( r \) is the rank of \( G \) and \( C_{G} \) its center. For the
particular case of \( G=SU(N) \), that corresponds to the symmetry breaking
\[
SU(N)\, \rightarrow \, U(1)^{N-1}\, \rightarrow \, Z_{N}.\]
For the first phase transition magnetic monopoles will appear. Then, in the
second phase transition magnetic flux tubes or strings (if \( C_{G} \) is non-trivial)
will appear and the monopoles will become confined. In order to produce this
symmetry breaking we shall look for vacuum solutions of the form 
\begin{eqnarray}
\phi _{1}^{\textrm{vac}} & = & a_{1}T_{+}\, ,\nonumber \\
\phi _{2}^{\textrm{vac}} & = & a_{2}T_{-}\, ,\label{3.3} \\
\phi _{3}^{\textrm{vac}} & = & a_{3}T_{3}\, ,\nonumber \\
W^{\textrm{vac}}_{\mu } & = & 0\, ,\nonumber 
\end{eqnarray}
where \( a_{1} \) and \( a_{2} \) are complex constants, \( a_{3} \) is a
real constant, and 
\begin{eqnarray*}
T_{3} & = & \delta \cdot H\, ,\, \, \, \, \, \delta \equiv \sum _{i=1}^{r}\frac{2\lambda _{i}}{\alpha _{i}^{2}}=\frac{1}{2}\sum _{\alpha >0}\frac{2\alpha }{\alpha ^{2}},\\
T_{\pm } & = & \sum _{i=1}^{r}\sqrt{c_{i}}E_{\pm \alpha _{i}},
\end{eqnarray*}
with \( \alpha _{i} \) and \( \lambda _{i} \) being simple roots and fundamental
weights, respectively, and 
\[
c_{i}\equiv \sum _{j=1}^{r}\left( K^{-1}\right) _{ij},\]
with \( K_{ij}=2\alpha _{i}\cdot \alpha _{j}/\alpha _{j}^{2} \) being the Cartan
matrix. The generators \( T_{3} \), \( T_{\pm } \) form the so called principal
\( SU(2) \) subalgebra of \( G \). The vacuum configuration \( \phi _{3}^{\textrm{vac}} \)
breaks \( G \) into \( U(1)^{\textrm{r}} \) and then \( \phi _{1}^{\textrm{vac}} \)
or \( \phi _{2}^{\textrm{vac}} \) breaks it further to \( C_{G} \). Let 
\[
\alpha _{i}^{\textrm{v}}\equiv \frac{2\alpha _{i}}{\alpha ^{2}_{i}}\, \, \, ,\, \, \, \, \, \lambda _{i}^{\textrm{v}}\equiv \frac{2\lambda _{i}}{\alpha _{i}^{2}},\]
be the simple coroots and fundamental coweights, respectively. Then using the
relations
\begin{eqnarray*}
\lambda _{j}^{\textrm{v}} & = & \alpha ^{\textrm{v}}_{i}\left( K^{-1}\right) _{ij},\\
\lambda _{i}^{\textrm{v}}\cdot \alpha _{j} & = & \delta _{ij},
\end{eqnarray*}
we obtain from the vacuum equations \( d_{a}=0=f_{sa} \), that
\begin{eqnarray*}
\left( a_{3}-\frac{\mu }{e}\right) a_{i} & = & 0\, \, ,\, \, \, \, \textrm{for }i=1,2\, ,\\
a_{1}a_{2} & = & \frac{\mu _{3}a_{3}}{e}\, ,\\
ma_{3} & = & \left| a_{2}\right| ^{2}-\left| a_{1}\right| ^{2}\, .
\end{eqnarray*}

Independently of the values of the mass parameters, this system always has the
trivial solution \( a_{1}=a_{2}=a_{3}=0 \), which corresponds to the vacuum
in which the \( G \) is unbroken. In \cite{k2003} the symmetry breakings produced
by the vacuum configuration given by Eq. (\ref{3.3}) were analyzed depending
on the values of mass parameters. We concluded that in the \( \mathcal{N}=4 \)
and \( \mathcal{N}=2^{*} \) theory (where \( \mu \neq 0 \)), the gauge group
\( G \) can be broken to \( U(1)^{r} \) which corresponds to the Coulomb phase.
Then, the gauge group can be further broken to \( C_{G} \), if we add to the
\( \mathcal{N}=2^{*} \) theory, a \( \mathcal{N}=1 \) or \( \mathcal{N}=0 \)
deformation (or both). Let us analyze each of these phases in the next sections.

\section{Coulomb phase}

In this phase \( G \) is broken to \( U(1)^{r} \) and there exist solitonic
monopole solutions. As we have seen, this phase can only occur for the \( \mathcal{N}=4 \)
and \( \mathcal{N}=2^{*} \) cases. That could happen, for example, for energy
scales in which one can consider \( \mu _{3}=0=m \). In this phase \( a_{1}=0=a_{2} \)
and \( a_{3}\neq 0 \). In principle \( a_{3} \) is an arbitrary non-vanishing
constant. However, we shall fix 
\[
a_{3}=\frac{\mu }{e}\]
in order to have the same value as in the Higgs phase. The vacuum solution \( \phi _{3}^{\textrm{vac}} \)
is the generator of a particular \( U(1) \) direction which we call \( U(1)_{\delta } \).
Since for any root \( \alpha  \), \( \delta \cdot \alpha \neq 0 \), we can
construct a monopole solution for each root \( \alpha  \). The associated monopole
magnetic charge is 
\begin{equation}
\label{4.2}
g\equiv \frac{1}{|\phi _{3}^{\textrm{vac}}|}\int dS_{i}\, \textrm{Re}\left( \phi ^{a}_{3}\right) B^{a}_{i}=\frac{2\pi }{e}\frac{\delta \cdot \alpha ^{\textrm{v}}}{|\delta |}.
\end{equation}
 Clearly \( g \) is equal to the monopole magnetic flux in the \( U(1)_{\delta } \)
direction, \( \Phi _{\textrm{mon}} \). Similarly one can define magnetic fluxes
\( \Phi ^{(i)}_{\textrm{mon}} \) associated to each \( U(1) \) factor of the
unbroken group \( U(1)^{r} \) which gives 
\begin{equation}
\label{4.3a}
\Phi _{\textrm{mon}}^{(i)}=\frac{2\pi }{e}\lambda ^{\textrm{v}}_{i}\cdot \alpha ^{\textrm{v}}.
\end{equation}

These are BPS monopoles with masses given by the central charge of the \( \mathcal{N}=2 \)
algebra \cite{witten olive, sw2}. For monopoles with vanishing fermion number,
their masses are \( M_{\textrm{mon}}=|g||\phi ^{\textrm{vac}}_{3}| \). Not
all of these monopoles are stable. The stable or fundamental are the ones with
lightest masses. For the present symmetry breaking, their masses are
\begin{equation}
\label{4.4}
M^{\textrm{L}}_{\textrm{mon}}=\frac{2\pi }{e|\delta |}|\phi _{3}^{\textrm{vac}}|.
\end{equation}
Note that, since \( G \) is completely broken to \( U(1)^{\textrm{r}} \),
differently from from the monopoles considered in the previous sections, here
the fundamental monopoles do not fill representations of a non-Abelian unbroken
group.

\section{Higgs or color superconducting phase}

In the Higgs or color superconducting phase, \( G \) is broken to its center
\( C_{G} \). That can happen when \( \mathcal{N}=2^{*} \) is broken by an
\( \mathcal{N}=1 \) or \( \mathcal{N}=0 \) deformation term (or both). In
this phase, the monopole chromomagnetic flux lines cannot spread out radially
over space. A phenomenon like that is expected to happen in the interior of
very dense neutron stars \cite{colorsupercondunting}. However, since 
\begin{equation}
\label{5.0}
\Pi _{1}\left( G/C_{G}\right) =C_{G},
\end{equation}
 if \( C_{G}=Z_{N} \), these flux lines can form topologically nontrivial \( Z_{N} \)
strings. Then, the monopoles of \( \mathcal{N}=2^{*} \) become confined in
this phase, as shown below.

The string tension bound given by Eq. (\ref{4.4d}) holds for \( \phi _{1} \)
and \( \phi _{2} \) in an arbitrary representation. In particular it holds
for the adjoint representation, which is the case we are considering here. Therefore,
since \( |\phi _{3}^{\textrm{vac}}|=\mu |\delta |/e \) in this phase, it results
that \cite{k2003} 

\begin{equation}
\label{5.4}
T\geq \frac{me}{2}\left| \phi ^{\textrm{vac}}_{3}\right| \left| \Phi _{\textrm{st}}\right| =\frac{m\mu }{2}\left| \delta \right| \left| \Phi _{\textrm{st}}\right| 
\end{equation}
 where, \( \Phi _{\textrm{st}} \) is the string flux, given by Eq. (\ref{4.3}).
The bound in Eq. (\ref{5.4}) holds for the BPS strings which satisfies the
equations \cite{k2003}
\begin{eqnarray}
B_{3a} & = & \mp d_{a,}\label{2.4a} \\
D_{\mp }\phi _{s} & = & 0,\label{2.4b} \\
f_{s} & = & 0,\label{2.4c} \\
E_{ia} & = & B_{1a}=B_{2a}=D_{0}\phi _{s}=D_{3}\phi _{s}=0,\label{2.4d} 
\end{eqnarray}

In order to have finite string tension, the string solution must satisfy the
vacuum equations asymptotically, which implies that 
\begin{eqnarray*}
\phi _{s}(\varphi ,\rho \rightarrow \infty ) & = & g(\varphi )\phi _{s}^{\textrm{vac}}g(\varphi )^{-1},\\
W_{I}(\varphi ,\rho \rightarrow \infty ) & = & g(\varphi )W_{I}^{\textrm{vac}}g(\varphi )^{-1}-\frac{1}{ie}\left( \partial _{I}g(\varphi )\right) g(\varphi )^{-1},
\end{eqnarray*}
 where \( \rho  \) is the radial coordinate and capital Latin letters \( I,J \)
denote the coordinates \( 1 \) and \( 2 \); \( \phi _{s}^{\textrm{vac}} \)
and \( W_{I}^{\textrm{vac}} \) are given by Eq. (\ref{3.3}) and \( g(\varphi )\in G \).
In order for the field configuration to be single valued, \( g(\varphi +2\pi )g(\varphi )^{-1}\in C_{G} \).
Considering 
\[
g(\varphi )=\exp i\varphi M\, ,\]
 then \( \exp 2\pi iM\in C_{G} \). That implies that \( M \) must be diagonalizable
and we shall consider that 
\[
M=\omega \cdot H.\]
Then, in order to \( \exp 2\pi i\omega \cdot H\in C_{G} \), 
\[
\omega =\sum _{i=1}^{r}l_{i}\lambda ^{\textrm{v}}_{i},\]
where \( l_{i} \) are integer numbers; that is, \( \omega  \) must be a vector
in the coweight lattice of \( G \), which has the fundamental coweights \( \lambda _{i}^{\textrm{v}} \)
as basis vectors. In principle, we could have other possibilities for \( M \)
which however we shall not discuss here. 

From this asymptotic configuration, in \cite{k2003} we construct a string anstaz
and obtained that 
\begin{equation}
\label{5.5}
\Phi _{\textrm{st}}=\frac{2\pi }{e}\frac{\delta \cdot \omega }{|\delta |}
\end{equation}

Similarly to the monopole, we can define string fluxes \( \Phi ^{(i)}_{\textrm{st}} \)
associated with the generators of each \( U(1) \) factor of \( U(1)^{r} \)
which results
\begin{equation}
\label{5.6}
\Phi ^{(i)}_{\textrm{st}}=\frac{2\pi }{e}\lambda ^{\textrm{v}}_{i}\cdot \omega \, .
\end{equation}
 Let us now check if the magnetic fluxes of the monopoles are compatible with
the ones of the strings. Since an arbitrary coroot \( \alpha ^{\textrm{v}} \)
can always be expanded in the coweight basis as \( \alpha ^{\textrm{v}}=\sum _{i=1}^{r}n_{i}\lambda _{i}^{\textrm{v}} \)
where \( n_{i} \) are integer numbers, one can conclude that the magnetic fluxes
(\ref{4.2}) or (\ref{4.3a}) of the monopoles can be expressed as an integer
linear combination of the string fluxes (\ref{5.5}) or (\ref{5.6}). Therefore,
in the Higgs phase, the monopole magnetic flux lines can no longer spread radially
over the space, since \( G \) is broken to the discrete group \( C_{G} \).
However, they can form one or more flux tubes or strings, and the monopoles
can become confined. In the next section, some concrete examples are given for
the case \( G=SU(3) \). We shall call this set of strings attached to a monopole
as confining strings. This set of confining strings must have total flux given
by Eq. (\ref{5.5}) or (\ref{5.6}) with \( \omega =\alpha ^{\textrm{v}} \).
That means that this set of confining magnetic strings belongs to the trivial
topological sector of \( \Pi _{1}(G/C_{G}) \) since \( \exp 2\pi i\alpha ^{\textrm{v}}\cdot H=1 \)
in \( G \). The fact that the set of confining strings must belong to the trivial
sector is consistent with the fact that the set is not topologically stable
and therefore can terminate at some point, like for the strings which appear
in the other type of symmetry breaking. Once more, it is important to stress
the fact that a string configuration belonging to the topological trivial sector
does not imply that its flux must vanish as we can see from Eq. (\ref{5.5}).
Again all these results are generalizations of some results for the \( Z_{2} \)
string of \( SU(2) \) Yang-Mills-Higgs theory. In the Higgs phase, string configurations
can in principle exist with flux \( 2\pi n/e \) for any integer \( n \), although
only the ones with \( n=\pm 1 \) are topologically stable. The ones with odd
\( n \) belong to the topologically nontrivial sector while the ones with even
\( n \) belong to the trivial sector. Therefore string configurations belonging
to the same topological sector do not have necessarily the same flux and therefore
are not related by (nonsingular) gauge transformations \cite{hk review, hk}.
As we mentioned before, the string configuration with \( n=2 \), belonging
to the trivial sector and which can be formed by two strings with \( n=1 \),
is the one which can terminate in the 't Hooft-Polyakov monopole with magnetic
charge \( g=4\pi /e \). In more algebraic terms one can say that this set of
integer numbers \( n \) forms the coweight lattice \( \Lambda _{\textrm{w}} \)
of \( SU(2) \), the subset of even numbers \( 2n \) form the \( SU(2) \)
coroot lattice \( \Lambda _{\textrm{r}} \), and the quotient \( \Lambda _{\textrm{w}}/\Lambda _{\textrm{r}}\simeq Z_{2} \)
corresponds to the center of \( SU(2) \). Therefore this quotient has two elements
which are represented by the cosets \( \Lambda _{\textrm{r}} \) and \( 1+\Lambda _{\textrm{r}} \).
Each coset corresponds to a string topological sector, with \( \Lambda _{\textrm{r}} \)
been the trivial one. 

In \cite{k2003}, this result was generalized for an arbitrary \( G \). Let
us for simplicity consider the case \( G=SU(N) \). Since \( SU(N) \) is simply
laced (i.e., \( \alpha ^{2}=2 \) for all roots \( \alpha  \)), we do not need
to distinguish between weights and coweights, roots and coroots. In this case,
the string topological sectors are given by 
\[
\Pi _{1}\left( SU(N)/Z_{N}\right) =Z_{N}\]
 and are associated with the \( N \) cosets 
\begin{equation}
\label{6.2}
\Lambda _{\textrm{r}}(SU(N))\, \, \textrm{and}\, \, \lambda _{i}+\Lambda _{\textrm{r}}(SU(N)),\, \, \, \, \, \, i=1,\, 2,\, ...\, ,N-1
\end{equation}
 where \( \lambda _{i} \) are the fundamental weights of \( SU(N) \) and \( \Lambda _{\textrm{r}}(SU(N)) \)
is the root lattice of \( SU(N) \). The coset \( \Lambda _{\textrm{r}}(SU(N)) \)
corresponds to the trivial topological sector. 

Since the confining string configuration linking a monopole to an antimonopole
belongs to the trivial topological sector, it can break when it has enough energy
to create a new monopole-antimonopole pair. As was done for the previous example
of monopole confinement, one can obtain a bound for the threshold length \( d^{\textrm{th}} \)for
the string breaking, using the relation
\begin{equation}
\label{5.9}
2M^{\textrm{L}}_{\textrm{mon}}=E^{\textrm{th}}=Td^{\textrm{th}}\geq \frac{me}{2}\left| \phi ^{\textrm{vac}}_{3}\right| \left| \Phi _{\textrm{st}}\right| d^{\textrm{th}}\, ,
\end{equation}
where \( E^{\textrm{th}} \) is the string threshold energy and \( M^{\textrm{L}}_{\textrm{mon}} \)
is the mass of the lightest monopoles, given by Eq. (\ref{4.4}). In the above
relation we used the string bound given by Eq. (\ref{5.4}) and did not consider
a possible energy term proportional to the inverse of the monopole distance,
known as the Lucher term. The modulus of the string flux, \( |\Phi _{\textrm{st}}| \),
must be equal to the modulus of the magnetic charges \( |g| \) of each confined
monopoles. Let us consider that \( |g|=2\pi |\delta \cdot \beta ^{\textrm{v}}|/|\delta | \)
with \( \beta ^{\textrm{v}} \) being an arbitrary coroot. Therefore one can
conclude from Eq. (\ref{5.9}), using Eq. (\ref{4.4}), that
\[
d^{\textrm{th}}\leq \frac{4}{me|\delta \cdot \beta ^{\textrm{v}}|}.\]

\section{Monopole confinement for \protect\( SU(3)\protect \) broken to \protect\( Z_{3}\protect \)}

Let us consider \( G=SU(3) \). We have seen that the magnetic lines of a given
monopole can form a set of flux tubes or strings. However, there are countless
different string configurations with this magnetic flux. It is not clear at
the moment which could be the preferable one. The most ``economical'' sets
would be the ones formed by a strings and an antistring as we shall see now. 

For \( SU(3) \), the different string topological sectors are associated with
the cosets 
\[
\Lambda _{\textrm{r}}(SU(3)),\, \, \lambda _{1}+\Lambda _{\textrm{r}}(SU(3))\, \textrm{and }\lambda _{2}+\Lambda _{\textrm{r}}(SU(3)).\]
One can, for example, construct string solutions associated with each of the
three weights \( \lambda _{1},\, \lambda _{1}-\alpha _{1},\, \lambda _{1}-\alpha _{1}-\alpha _{2} \)
of the three dimensional fundamental representation. Since all of them belong
to the coset \( \lambda _{1}+\Lambda _{\textrm{r}}(SU(3)) \), these string
solutions belong to the same topological sector. However, one can observe from
Eq. (\ref{5.5}) that they do not have the same flux \( \Phi _{\textrm{st}} \),
similarly to the \( Z_{2} \) strings of \( SU(2) \) theory. Therefore these
string solutions are \textit{not} related by gauge transformations since \( \Phi _{\textrm{st}} \)
is gauge invariant. One can construct the corresponding antistring solutions
associated with the negative of these weights, which form the complex-conjugated
representation \( \overline{3} \) and which belong to the coset \( \lambda _{2}+\Lambda _{\textrm{r}} \).
The magnetic fluxes of the monopoles associated with the six non-vanishing roots
of \( SU(3) \) can easily be written using these strings in the following way:
for the monopole \( \alpha _{1} \) we can attach the strings \( \lambda _{1} \)
and \( -\lambda _{1}+\alpha _{1} \). For the monopole \( \alpha _{2} \) we
can attach strings \( \lambda _{1}-\alpha _{1} \) and \( -\lambda _{1}+\alpha _{1}+\alpha _{2} \).
For the monopole \( \alpha _{1}+\alpha _{2} \) we can attach the strings \( \lambda _{1} \)
and \( -\lambda _{1}+\alpha _{1}+\alpha _{2} \). And similarly for the other
three monopoles associated with the negative roots, just changing the signs.
The remaining three combinations of strings and antistring have vanishing fluxes
\( \Phi ^{(i)}_{\textrm{st}} \).

\vspace{0.3cm}
\begin{figure}
\begin{center}
\resizebox*{0.97\columnwidth}{!}{\rotatebox{270}{\includegraphics{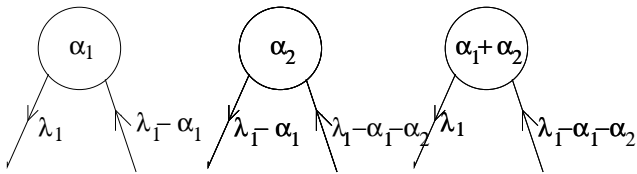}}}  
\caption{Strings attached to monopoles for $G=SU(3)$.}
\end{center}
\end{figure}
\vspace{0.3cm}

One could draw the above set of strings attached to monopoles as shown in Fig.1,
where the circles represent the monopoles and the arrows are the string flux
\( \Phi ^{(i)}_{\textrm{st}} \). We represented the strings associated with
weights in the fundamental representation by an arrow going out of the monopole
and for the antistrings we reversed the sense of the arrow and simultaneously
changed the sign of the weight. Then, in addition to the monopole-antimonopole
pairs one could also conjecture about the formation of a confined system with
the monopoles \( \alpha _{1} \), \( \alpha _{2} \) and \( -\alpha _{1}-\alpha _{2} \)
as shown in Fig. 2. Note that since these monopoles are not expected to fill
the three dimensional fundamental representation of \( SU(3) \), that system
is not exactly like a baryon. With this configuration of monopoles with strings
attached, one could also think of putting one string in the north pole and the
on the other in the south pole, forming a configuration similar to the bead
described in \cite{hk}. One can easily extend this construction of strings
attached to monopoles and monopole confined system to the \( SU(N) \) case\cite{k2003}. 

\vspace{0.3cm}
\begin{figure}
\begin{center}
\resizebox*{0.85\columnwidth}{!}{\rotatebox{270}{\includegraphics{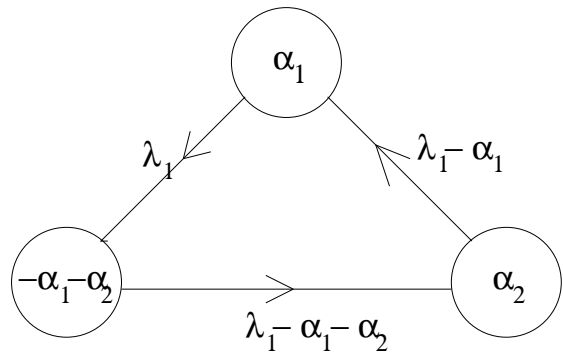}}}  
\caption{Confined system of three monopoles for $G=SU(3)$.}
\end{center}
\end{figure}
\vspace{0.3cm}

\section{String tension and Casimir scaling law}

The string tension is one of the main quantities to be determined in quark confinement
in QCD. In these last 20 years quite a lot of work has been done trying to determine
this quantity. There are mainly two conjectures for the string tension: the
``Casimir scaling law'' \cite{casimir} and the ``sine law'' \cite{douglas shenker}.
In these two conjectures the gauge group \( G=SU(N) \) is considered and a
string in the representation associated with the fundamental weight \( \lambda _{k} \)
which can be obtained by the antisymmetric tensor product of \( k \) fundamental
representations associated with \( \lambda _{1} \).  For the Casimir scaling
conjecture, the string tension should satisfy 
\begin{equation}
\label{7.1}
T_{k}=T_{1}\frac{k(N-k)}{N-1},\, \, \, \, \, \, k=1,\, 2,\, ...,\, N-1,
\end{equation}
 where \( T_{1} \) would be the string tension in the \( \lambda _{1} \) fundamental
representation. On the other hand, for the sine law conjecture, 
\[
T_{k}=T_{1}\frac{\sin \left( \frac{\pi k}{N}\right) }{\sin \left( \frac{\pi }{N}\right) },\, \, \, \, \, \, k=1,\, 2,\, ...,\, N-1.\]
 All these conjectures are concerned with the chromoelectric strings. However,
as we mentioned in the introduction, one expects that chromomagnetic strings
could be related to chromoelectric strings by a duality transformation. Therefore
one could ask if the tensions of our chromomagnetic strings satisfy one of the
two conjectures. 

For the case \( G=SU(N) \), for a string associated with the weight \( \omega  \),
such that 
\[
\omega =\lambda _{k}-\beta _{\omega },\]
where \( \lambda _{k} \) is a fundamental weight of \( SU(N) \) and \( \beta _{\omega }\in \Lambda _{\textrm{r}}(SU(N)) \),
the string tension bound, given by Eq. (\ref{5.4}), can be written as 
\begin{equation}
\label{7.2}
T_{\lambda _{k}-\beta _{\omega }}\geq \frac{m\mu \pi }{e}\left| \frac{1}{2}\left[ C(\lambda _{k})-\lambda _{k}\cdot \lambda _{k}\right] -\delta \cdot \beta _{\omega }\right| ,
\end{equation}
 where
\[
C(\lambda _{k})=\lambda _{k}\cdot \left( \lambda _{k}+2\delta \right) \]
 is the quadratic Casimir associated with the fundamental representation \( \lambda _{k} \).
That expression can be also written as 
\begin{equation}
\label{7.3}
T_{\lambda _{k}-\beta _{\omega }}\geq \frac{m\mu \pi }{e}\left| \frac{1}{2}\left( \frac{\left( N-1\right) ^{2}}{2N}\frac{k\left( N-k\right) }{N-1}\right) -\delta \cdot \beta _{\omega }\right| 
\end{equation}
 Therefore the first term on the right-hand-side of this inequality or, equivalently,
the BPS string tension associated with \( \omega =\lambda _{k} \) can be written
as 
\begin{equation}
\label{7.4}
T^{{\scriptsize \textrm{ BPS}}}_{\lambda _{k}}=T^{{\scriptsize \textrm{ BPS}}}_{\lambda _{1}}\frac{k\left( N-k\right) }{N-1},\, \, k=1,\, 2,\, ...,\, N-1,
\end{equation}
where 
\[
T_{\lambda _{1}}^{{\scriptsize \textrm{ BPS}}}=\frac{m\mu \pi }{2e}\frac{\left( N-1\right) ^{2}}{2N}\]
 is the BPS string tension associated with \( \omega =\lambda _{1} \). Hence
we explicitly showed that the BPS string tensions associated with an arbitrary
\( SU(N) \) fundamental weight \( \lambda _{k} \) satisfy the Casimir scaling
conjecture, given by Eq. (\ref{7.1}). 

\vskip 0.2 in \noindent \textbf{}\textbf{\large Acknowledgments} \textbf{}

\noindent

I would like to thank the organizes of this meeting for the invitation to present
this seminar and FAPERJ for financial support.

\end{document}